\documentclass[article,twocolumn]{revtex4}
\usepackage{
  amsmath,
  amssymb,
  amsthm,
  bm,
  dcolumn,
  fancyvrb,
  framed,
  graphicx,
  hyperref,
  mathtools,
  morefloats,
  textcomp,
  url,
}
\usepackage[normalem]{ulem}
\usepackage[caption=false]{subfig}

\begin{document}
\title{Design optimization of passively mode-locked semiconductor lasers with intracavity grating spectral filters}

\author{Finbarr O'Callaghan}
\affiliation{Tyndall National Institute, Lee Maltings, University College Cork, Cork, Ireland}
\author{David Bitauld}
\thanks{Current address: Nokia Research Center, 21 J J Thomson Avenue, Cambridge CB3 0FA, UK}
\affiliation{Tyndall National Institute, Lee Maltings, University College Cork, Cork, Ireland}
\author{Stephen O'Brien}
\affiliation{Tyndall National Institute, Lee Maltings, University College Cork, Cork, Ireland}

\begin{abstract}
We consider design optimization of passively mode-locked two-section semiconductor lasers that incorporate intracavity 
grating spectral filters. Our goal is to develop a method for finding the optimal wavelength location for the filter 
in order to maximize the region of stable mode-locking as a function of drive current and reverse bias in the absorber 
section. In order to account for material dispersion in the two sections of the laser, we use analytic approximations 
for the gain and absorption as a function of carrier density and frequency. Fits to measured gain and absorption 
curves then provide inputs for numerical simulations based on a large signal accurate delay-differential model of the 
mode-locked laser. We show how a unique set of model parameters for each value of the drive current and reverse bias voltage 
can be selected based on the variation of the net gain along branches of steady-state solutions of the model. We demonstrate 
the validity of this approach by demonstrating qualitative agreement between numerical simulations and the measured 
current-voltage phase-space of a two-section Fabry-Perot laser. We then show how to adapt this method to determine an 
optimum location for the spectral filter in a notional device with the same material composition, based on the targeted 
locking range, and accounting for the modal selectivity of the filter. 
\end{abstract}

\maketitle

\section{Introduction}

Semiconductor mode-locked lasers have the potential to address a great number of applications 
in advanced telecommunications and signal processing \cite{avrutin_00, kaiser_07, quinlan_09, 
msallem_11}. Because many of these applications place stringent requirements on the laser source, 
a series of design innovations have been suggested that can enhance the timing and phase-noise 
performance of semiconductor mode-locked [ML] lasers. In particular, devices that incorporate 
intracavity spectral filters and pulse shapers have seen significant progress. Recent examples 
include integration of a Mach-Zehnder interferometer for flattening of the gain spectrum 
\cite{parker_12}, integration of arrayed waveguide gratings and phase modulators for pulse 
shaping \cite{heck_08}, and harmonic mode-locking of extended cavity devices with integrated 
ring resonator filters for phase-noise reduction \cite{srinivasan_14}. In parallel with these 
developments, a further series of interesting experimental techniques and devices have been 
demonstrated that exploit the very large quality factors of whispering gallery mode microresonators 
in order to generate low phase-noise optical frequency combs and stable sources of ultrafast pulses. 
Many of these innovations are based on principles of multiwavelength excitation of parametric 
processes \cite{strekalov_09, papp_14} or on coupling of conventional lasing and nonlinear 
resonators \cite{peccianti_12, johnson_14}. 

In recent work we have demonstrated a number of two-section Fabry-Perot lasers with engineered 
spectra defined by an intracavity grating spectral filter \cite{obrien_10, bitauld_10}. The grating 
filter in these devices is designed to select a finite number of predetermined lasing modes so that 
precise tailoring of the comb line spectrum is possible. These filter designs can also be 
adapted for integrated lightwave circuits based on open grating resonators to provide on-chip 
sources of tailored and phase-locked lasing modes \cite{obrien_14}.

In conventional two-section semiconductor lasers, the peak emission wavelength can vary strongly as 
the current and the reverse bias applied to the short section are varied \cite{stolarz_11}. Although 
we have succeeded in mode-locking a variety of devices that incorporated grating spectral filters  
\cite{bitauld_11a}, we have found that the extent of the stable mode-locking region was very limited 
compared to Fabry-Perot [FP] lasers with the same material composition. The ability to adjust the laser 
drive parameters while remaining in a mode-locked state is an important requirement for applications, 
as this will enable tuning of the spectral profile and comb line frequencies. The latter property is 
of great practical importance, as it will facilitate locking of a comb to an external pump beam, or to 
an extended underlying or external cavity. Here we develop a design strategy for optimising the 
mode-locking range of two-section semiconductor lasers with intracavity grating spectral filters. 

In order to optimize the mode-locking range of these devices for applications, efficient dynamical 
models are crucial for understanding the structure of the mode-locking region in the phase-space of drive 
parameters. In general, frequency domain models extend the rate equation description of a two-section semiconductor 
laser \cite{dubbeldam_99, ocallaghan_14} to include phase-sensitive modal interactions \cite{lau_90, avrutin_03}. 
In this case, the mode-locked state can be described as a mutually injection locked steady-state with zero net 
group velocity dispersion. Although these models are attractive and convenient given the natural modal picture 
that follows from our design approach \cite{ocallaghan_14}, current frequency domain models are valid for small 
gain and loss. In addition, because these models describe the ML state as a steady-state rather than a steady 
\textit{periodic} state, these models are small signal models of the mode-locked laser \cite{avrutin_book}. 
They therefore may not provide a complete and accurate picture of the various dynamical processes that can 
destabilize the ML state \cite{vladimirov_05, rachinskii_06}. Fully distributed time domain simulations that 
describe the spatio-temporal dynamics of the carriers and the propagating fields provide a quantitative picture 
of the dynamics in semiconductor ML lasers \cite{bischoff_95, mulet_06, javaloyes_10}. A lumped element time 
domain model has also been developed that can describe the large gain and losses and strong saturation of the 
absorption that are typical in a semiconductor laser \cite{vladimirov_04}. We will employ this delay differential 
model for our numerical simulations, which has advantages of computational efficiency and potential for analytic 
analysis \cite{vladimirov_05}.

Here we first present experimental results that illustrate the complex dynamical phase-space structure that 
is typical of a two-section FP laser. We  map regions of self-pulsations, stable mode-locking, and 
Q-switched mode-locking as a function of injected current and reverse bias applied to the short section 
of the laser. We also present optical spectra showing how the peak emission can vary strongly and exhibit large 
shifts over relatively small changes in the drive parameters. We then provide measurements from a device with 
a tailored spectrum, which show an unexpected variation of the net gain at the location of the 
spectral filter.

Our design method requires us to describe the frequency dispersion of gain and saturable absorption and to 
capture their effects on the dynamics and stability of mode-locked states of these devices. To calibrate our model 
parameters, we fit an approximate analytic function to the measured modal gain and to the modal absorption as a 
function of reverse bias. To reproduce the measured phase-space data, we propose a direct method to choose 
a wavelength reference based on the variation of the net gain along steady-state solutions of the delay 
differential equation [DDE] model. This approach leads to good qualitative agreement between the DDE model 
and the results of our experiments.

We next consider the problem of finding the optimal wavelength location for an intracavity spectral filter. Key 
properties of interest are the stability of the ML spectrum as defined by the grating filter and tunability of the 
ML state. Because the ML states of our devices are characterised by relatively narrow spectral bandwidths, we consider 
a quasi-static limit, and we again use steady-state solutions to estimate the variation of the net gain at detuned 
wavelengths over a defined region in the phase-space of drive parameters. Based on measured results for the selectivity 
of typical grating filters and the structure of the simulated mode-locking region, our results indicate that we can 
expect stable mode-locking in optimized devices over a large range of voltages and currents. We conclude by discussing 
possible improvements to the proposed method and prospects for future work. 

\section{Experimental measurements of the impact of material dispersion on passive mode-locking 
of two-section semiconductor lasers}

In this section we present experimental measurements taken from two-section indium phosphide-based ridge-waveguide FP 
lasers. The devices are high-reflection [HR] coated, with quantum-well active regions and the absorber section placed 
adjacent to the HR mirror. The first device is a plain FP laser of length 545 $\mu$m, with a saturable absorber section of 
length 30 $\mu$m, while the second is a device of length 875 $\mu$m, with a saturable absorber section of length 60 $\mu$m. 
The second device includes an intracavity grating spectral filter defined by etched features in the laser ridge-waveguide. 

\begin{figure}[t]
\centering
  \subfloat{\includegraphics[trim=0.5cm 0cm 1cm 0.5cm, clip=true, width=0.62\columnwidth,height=11em]{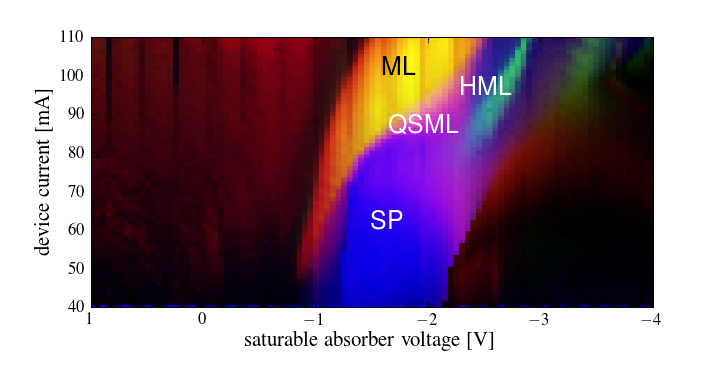}}
  \subfloat{\includegraphics[trim=0.5cm 0cm 0cm 0.75cm, clip=true, width=0.425\columnwidth,height=11em]{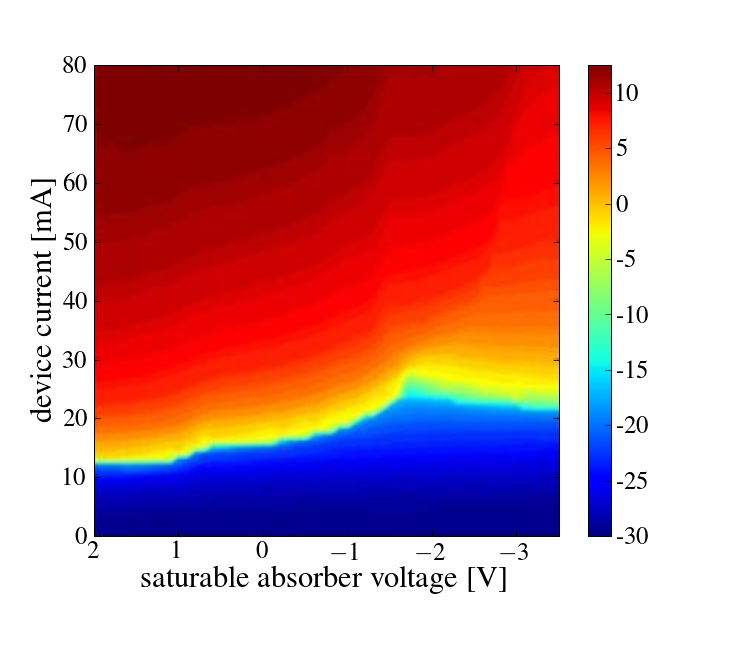}} \\
  \subfloat{\includegraphics[trim=1cm 8cm 2cm 2cm, clip=true, width=0.98\columnwidth,height=12em]{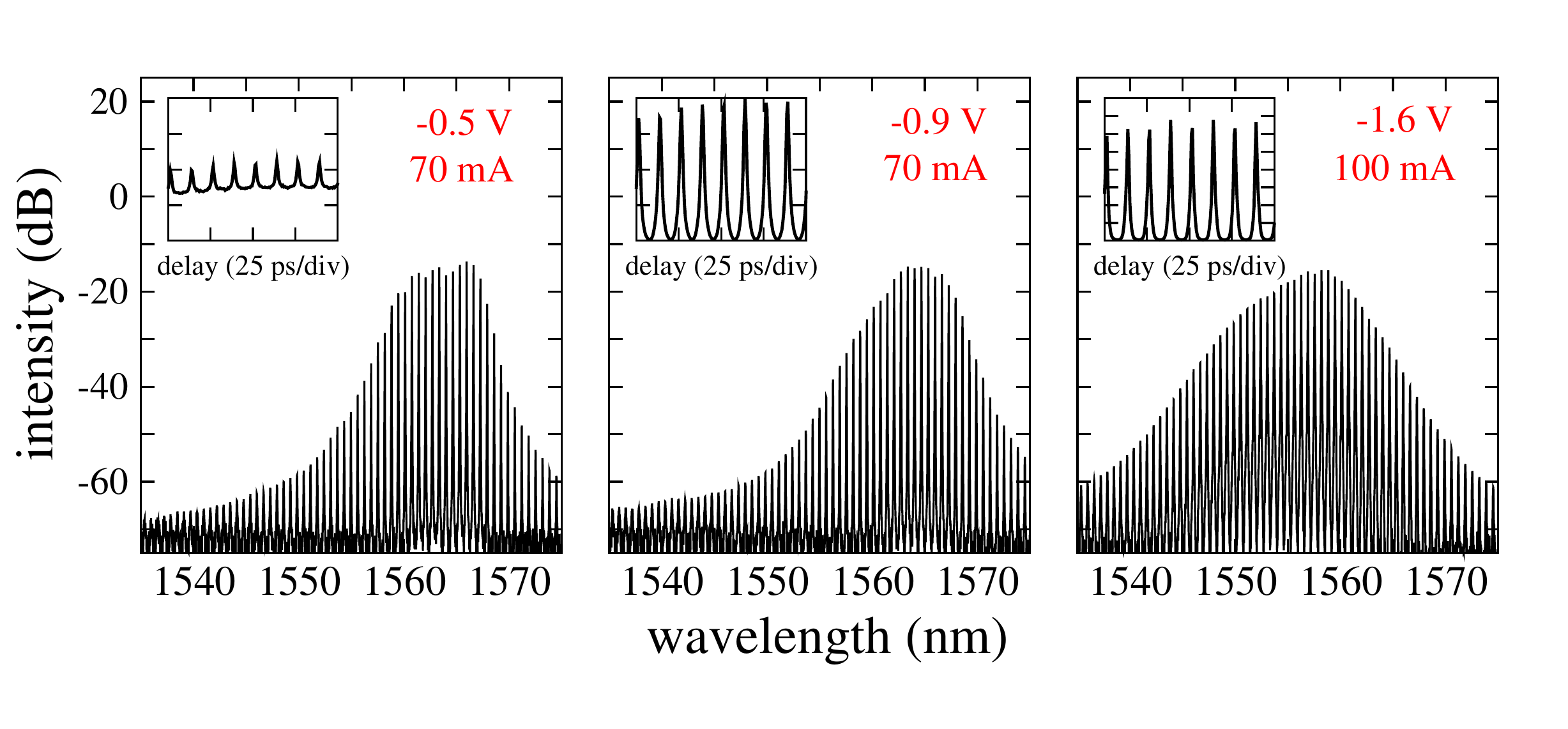}}
  \caption{\label{fig1}  Upper panel: Measured phase space data (left) and power output (right) of a two-section plain Fabry-Perot 
laser. Labeled phase-space regions are SP (self-pulsation), ML (mode-locking), QSML (Q-switched mode-locking), and HML 
(harmonic mode-locking). Lower panel: Optical spectra and intensity autocorrelation measurements for a series of drive parameters 
as indicated.}
\end{figure}

For the plain FP device the measured threshold current with a single current density over the whole device length 
[single section FP] was 12.5 mA. A measured phase-space map of the dynamical states of the device as the device current 
and reverse bias voltage applied to the short section are varied is shown in the upper panel of Fig. \ref{fig1}. One can 
see that a large region of self-pulsations [SP] is found near to threshold, and that these tend to evolve into mode-locked 
[ML] operation through a region of Q-switched mode-locking [QSML]. The current range where QSML is found becomes wider as 
the reverse bias is increased, and harmonic mode-locking [HML] is also found in a small region for large negative values of 
the absorber bias. A map of the power output of the device is also shown in the upper panel of Fig. \ref{fig1}. This shows 
an increasing threshold current until approximately -1.5 V, where an anomalous decrease is found for increasing negative bias. 

In the lower panel of Fig. \ref{fig1} we have included optical spectra and the corresponding intensity 
autocorrelation measurements at three points in the phase space of Fig. \ref{fig1}. We observe incomplete 
mode-locking at small values of the reverse bias around -0.5 V. Well developed mode-locking of the FP device is 
found near a device current of 70 mA, and a reverse bias voltage of -1.0 V. In this region the device generates 
pulses of approximately 1.7 ps duration at a repetition rate of 80 GHz. These pulses, which are not transform 
limited, become shorter as the reverse bias is increased, reaching a minimum duration of approximately 1.5 
ps near -1.5 V. Note that the shift of the peak emission wavelength with increasing reverse bias is initially 
towards longer wavelengths and that this trend is reversed for larger reverse bias voltages. This reversal also 
occurs in the region of reverse bias voltage values where the threshold variation with increasing bias becomes 
anomalous. 

\begin{figure}[t]
\centering
\includegraphics[trim=0.5cm 0cm 1cm 0cm, clip=true,width=0.95\columnwidth,height=15em]{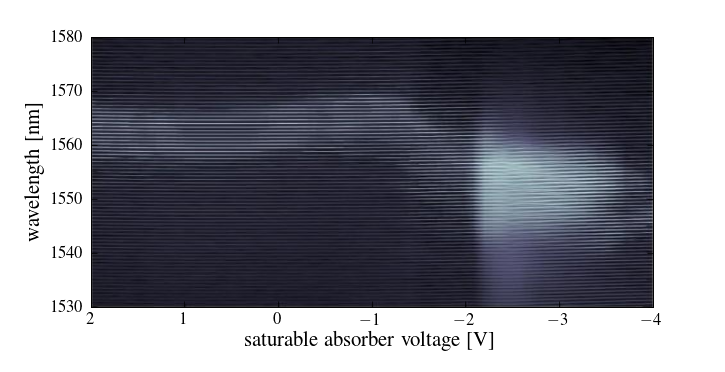}
\caption{\label{fig2} Optical spectrum of the two-section plain Fabry-Perot laser as the reverse bias voltage 
is varied. The current in the gain section is 100 mA. }
\end{figure}

Fig. \ref{fig2} illustrates the dramatic shifts in the peak emission wavelength that occur in the region of anomalous 
threshold variation. Here we have shown the optical spectrum of the device as a function of saturable absorber voltage 
at a fixed device current in the long section of 100 mA. At reverse bias voltages up to approximately -1.5 V, one can 
see that the peak emission wavelength is relatively constant, with a small drift towards long wavelength. Beyond this 
region, with increasing reverse bias, we observe a much more rapid shift of the peak emission towards shorter wavelengths. 

\begin{figure}[t]
\centering
\includegraphics[trim=0.5cm 1cm 2cm 2cm, clip=true, width=0.98\columnwidth]{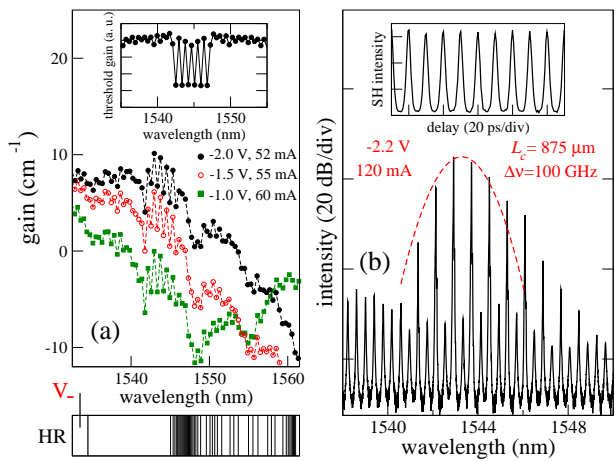}
\caption{\label{fig3} (a) Net gain measured below threshold for the drive parameters indicated. The inset 
shows a calculation of the modal threshold gain. The lower panel is a device schematic showing the two-section
geometry and the construction of the grating spectral filter. The device length is 875 $\mu$m. (b) Mode-locked 
spectrum of the device. The inset shows an intensity autocorrelation measurement. The dashed line is the implied 
spectral bandwidth derived from the autocorrelation measurement.}
\end{figure}

Experimental data from the second device of length 875 $\mu$m are presented in Fig. \ref{fig3} \cite{bitauld_10}. 
In this example the filter was designed to select a comb of six primary modes with 100 GHz spacing at a wavelength
of 1545 nm. Net gain measurements near threshold made using the Hakki-Paoli technique are shown in Fig. \ref{fig3} (a). 
These data show that the net gain is an \textit{increasing} function of reverse bias near the location of the 
spectral filter for reverse bias voltages larger than -1.0 V. Note also that Fig. \ref{fig3} (a)
provides us with a measure of the filter selectivity that can be achieved with etched features in devices such as 
these. In the device shown, the selectivity is somewhat non-uniform but it reaches a maximum of approximately 2 
$\mbox{ cm}^{-1}$. We found that the device of Fig. \ref{fig3} mode-locked near a drive current of 120 mA and reverse 
bias voltage of -2.2 V to obtain a transform limited 100 GHz pulse train with 2 ps pulse duration [Fig. \ref{fig3} (b)]. 
The timing jitter of the pulse train and the phase noise of the individual comb lines also showed significant 
improvement as compared to an equivalent plain FP laser \cite{bitauld_11a}, which we attribute to the beneficial 
effects of thinning of the optical spectrum. The optical spectrum also shows strong suppression of cavity modes
between the primary comb lines, which could otherwise be a source of so-called supermode noise \cite{quinlan_09, 
srinivasan_14}. However, although we obtained a high quality mode-locking spectrum, the stable ML region was far 
too small to allow significant tunability of the comb spectrum. 

The measurements of Fig. \ref{fig2} and Fig. \ref{fig3} (a) suggest that the limited mode-locking range of these 
devices was due to the complex dispersion of the net modal gain at the location of the spectral filter. In order to 
address this issue, we have developed a general method for determining the optimal wavelength location for a spectral 
filter that can lead to a large ML region and associated tunability in the parameter region where the variation of the 
peak wavelength of the net gain with drive parameters is less pronounced. We have noted that a spectral filter will 
constrain the active modes of a device so that the carrier wave frequency in the mode-locked state cannot adjust 
as in the FP. Although a quantitative optimization method would explicitly account for the action of the spectral 
filter, we will illustrate the application of the method by applying it to a notional device with the same material 
composition as the two-section FP laser of Fig. \ref{fig1}. However, the method is general, and it could be 
adapted to models of the two-section device that account for the filter explicitly. We also argue that our results 
here can be regarded a qualitatively accurate, as we expect model-locking of a narrow-band comb defined by an 
intracavity spectral filter will not be more difficult than a plain FP laser, provided that the additional modal 
dispersion due to the filter is limited.  

\section{Model parameter extraction based on an analytical model of the semiconductor modal 
gain and absorption}

In this section we describe the analytic model of the modal gain and absorption that we use to fit measured data.
Before we proceed, it is instructive to first introduce the well-known rate equation description of a laser with 
a saturable absorber [LSA model]. By considering these equations, which are also referred to as the Yamada model 
\cite{yamada_93}, we can introduce the various model parameters that we must obtain in their physical context, 
and we may also motivate the scaling of the dynamical variables that we will employ in our simulations.

Although the LSA model cannot describe mode-locked states of the device, it is the simplest model that can 
describe phenomena such as self-pulsations and the appearance of a bistable region at threshold in devices with 
saturable absorbers \cite{dubbeldam_99}. These equations make a total field approximation to describe the 
intensity of the FP laser, and the saturable absorber section of the laser is described as an unpumped region, 
with an unsaturated loss determined by the applied voltage. In physical units the LSA model reads 
\begin{eqnarray}\label{lsa_sm}
& \dot{S} = [(1 - \rho)G_m(N_{g}) + \rho Q_m(N_{q}) - \gamma] S \nonumber \\
& \dot{N}_{g} = j - \frac{N_{g}}{\tau_s} - G_m(N_{g}) S \\
& \dot{N}_{q} = \:\:\:\: - \frac{N_{q}}{\tau_q} - Q_m(N_{q}) S \nonumber
\end{eqnarray}
Here $S$ is the average photon density, $G_m$ is the modal gain and $Q_m$ is the modal absorption. 
$N_{g}$ and $N_{q}$ are the carrier densities in the gain and absorber sections, respectively. The total 
field losses are $\gamma = \alpha_{\mbox{\tiny m}} + \alpha_{\mbox{\tiny int}}$, where $\alpha_{\mbox{\tiny m}}$ are 
the mirror losses, and $\alpha_{\mbox{\tiny int}}$ are the internal losses of the device. The current density in the 
gain section is $j$, while the carrier lifetimes in the gain and absorber sections are $\tau_s$ and 
$\tau_q$ respectively. 

To proceed we must establish the dependence of the gain and absorption in the device 
as a function of carrier density and frequency. Reference \cite{balle_98} provides a convenient
model that can be fitted to measured gain and absorption obtained using the Hakki-Paoli method. This 
work developed an analytical expression for the susceptibility of a quantum-well semiconductor 
material at low-temperature and has been used extensively to model the gain and saturable absorption 
in simulations of free-running and mode-locked semiconductor lasers \cite{javaloyes_10, stolarz_11}. 
While the model results cannot be regarded as quantitative, they nevertheless provide a qualitative 
description of the material susceptibility that can be used to obtain physically appropriate 
parameters for dynamical simulations. 

The result for the modal gain is  \cite{balle_98}
\begin{eqnarray}\label{balle_chi-g}
G_m(\lambda, D_g) = G_0 (\tan^{-1}[u] - 2 \tan^{-1}[u - D_g ] - \pi/2 ), \nonumber
\end{eqnarray}
where $\lambda$ is the wavelength, $D_g$ is the carrier density normalised to the transparency value, 
$G_0$ is the material gain coefficient, and 
\begin{eqnarray}
u = \frac{2 \pi c }{\gamma_p^G}  \left(\frac{1}{\lambda} - \frac{1}{\lambda_{bg}^{G}} \right) + \sigma D_g^{1/3}. \nonumber 
\end{eqnarray}
Here, $\lambda_{bg}^{G}$ is the nominal transition wavelength, $\sigma$ describes the bandgap shrinkage 
with increasing carrier density, and $\gamma_p^G$ is the linewidth of the optical transition. We use a different 
carrier density variable in the expressions for the modal gain and absorption in Eqns \ref{lsa_sm} as we will 
choose to normalise the model variables with respect to the differential gain, rather than the transparency 
value of the carrier density. 

\begin{figure}[t]
\centering
\includegraphics[trim=0.5cm 1cm 1.5cm 2cm, clip=true, width=0.975\columnwidth]{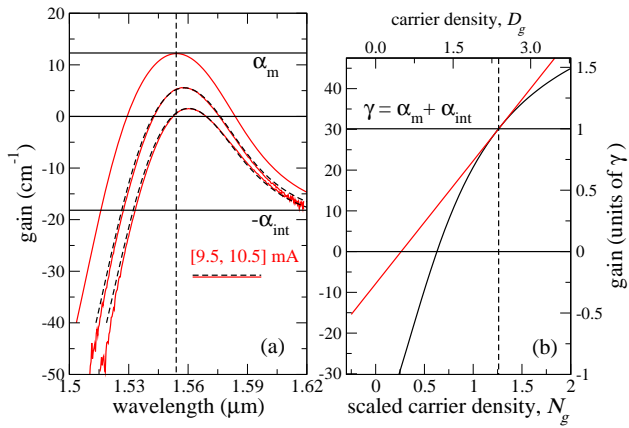}
\caption{\label{fig4} (a) Measured modal gain of the Fabry-Perot laser [dashed lines] for two  
values of the device current as indicated. Solid lines are fits to these data and a plot of the gain curve 
at threshold defined by the fitted gain function. The mirror losses are $\alpha_{\mbox{\tiny m}}$ and the background losses 
are $\alpha_{\mbox{\tiny int}}$. (b) Variation of the fitted modal gain as a function of carrier density at a 
fixed wavelength $\lambda_0$. This wavelength [1554 nm] is at the location of the peak gain at threshold for 
the Fabry-Perot laser indicated by the vertical dashed line in (a). The threshold modal gain, $\gamma$, is 
equal to the sum of the mirror and internal losses in the device. The differential gain at threshold is 
defined to be equal to one in normalised units. }
\end{figure}

We now fit this expression to the modal gain of the active material, $G_m$, which we determine using the
Hakki-Paoli technique. For these measurements a single current density over the full length of the device 
was maintained. The negative offset at long-wavelength gives an estimate of the internal losses, which are 
assumed to be wavelength independent and uniformly distributed over the device length. The results of the 
fit that we obtained for two values of the device current are shown in Fig. \ref{fig4} (a). The mirror 
losses of the device are $\alpha_{\mbox{\tiny m}} = 12.2 \mbox{ cm}^{-1}$. We find that the internal losses, 
$\alpha_{\mbox{\tiny int}} = 18.0 \mbox{ cm}^{-1}$, as indicated in Fig. \ref{fig4} (a).

From the resulting fit to the data, we determine the value of the carrier density at threshold for 
the single section FP laser. Fixing the wavelength at the location of the gain  peak at threshold, we  
determine the differential gain at threshold for the FP laser. We then define the threshold gain and 
differential gain at the gain peak to be equal to 1 in normalized units. This scale then defines our 
normalized carrier density variable, $N_g$. We will also define the scaled current density, $j_s$, where 
the threshold current density in the single section FP laser is defined to be equal to 1 in normalised units.  
The variation of the modal gain at the wavelength of the gain peak at threshold as a function of carrier 
density is shown in Fig. \ref{fig4} (b). Here have plotted the data in physical units and in normalized units. 
The value of the differential gain at threshold is $g_{g} =  16.0 \mbox{ cm}^{-1} [\equiv 1]$. A table of fitting 
parameters for the susceptibility function is presented in table \ref{t1}. The values of $D_g$ and $N_g$ at 
threshold are 2.38 and 1.26 respectively. 

\begin{table}\renewcommand{\arraystretch}{1.5}
\begin{center}
\begin{tabular}{cccccc} 	\hline
$\lambda_{bg}^{G}$ & $ G_0 $  & $\gamma_p^{G}$ & $\sigma$ & $\lambda_{0}$ & $g_{g}$ \\ \hline
 [nm] & [$\mbox{cm}^{-1}$] & [$10^{13}\mbox{ s}^{-1}$]  & & [nm] & [$\mbox{cm}^{-1}$] \\ \hline
1550.0 & 27.0 & $2.3$   & 0.73 & 1554.0 & 16.0 \\  \hline
\end{tabular}
\caption{ \label{t1} Fitting and derived parameters for the gain function}
\end{center}
\end{table} 

To determine the modal absorption in the reverse bias section, $Q_m$, we employ a differential Hakki-Paoli 
technique. The modal absorption is found from the difference between the 
modal gain measured with a uniform current density over the whole device length and the modal gain, $G_m^{'}$, 
measured with the same current density applied to the long section and a given voltage applied to the short 
section. The relationship is  \cite{stolarz_11, scheibenzuber_11}
\begin{displaymath}
\rho Q_m(V, \lambda) = G_m^{'}(V, j, \lambda) + \alpha_{\mbox{\tiny int}} - (1 - \rho)G_m(j, \lambda).
\end{displaymath}

To fit the derived absorption curve we use the expression
\begin{eqnarray}
&& Q_m (\lambda, D_q) + \alpha_{\mbox{\tiny abs}} = \nonumber \\
&& \:\:\:\:\: Q_0[\mbox{\small V}]( \tan^{-1}[ u ]  - 2 \tan^{-1}[ u - D_q ] - \pi/2 ), \nonumber
\end{eqnarray}
where $D_q$ is the carrier density normalised to the transparency value as before, the material absorption
coefficient, $Q_0[\mbox{\small V}]$, is now a function of voltage, and 
\begin{eqnarray}
u = \frac{2 \pi c }{\gamma_p^Q} \left(\frac{1}{\lambda} - 
\frac{1}{\lambda_{bg}^{Q}[\mbox{\small V}]} \right). \nonumber 
\end{eqnarray}
Here we have neglected the $\sigma$ parameter as the carrier density is taken to be much smaller in the 
reversed biased section. However, in order to obtain a reasonable fit to the data we have had to introduce
a voltage dependent bandgap wavelength $\lambda_{bg}^{Q}[\mbox{\small V}]$. We have also included an additional 
carrier density and voltage independent contribution to the absorption in the absorber section, 
$\alpha_{\mbox{\tiny abs}}$. This was necessary because, while the above expression can fit the absorption function 
accurately near the bandedge, we found that the derived carrier density was close to the transparency value, 
which led to an unphysically large value for the differential absorption of the material. We therefore included 
a small offset, $\alpha_{\mbox{\tiny abs}} = 2.5 \mbox{ cm}^{-1}$, in order to obtain accurate 
fits of the function with small values of the carrier density appropriate for a reverse biased section. Note 
also that the transition linewidth in the absorber section was taken to be $\gamma_p^Q = 0.85 \times 10^{13} 
\mbox{ s}^{-1}$. This change in the value of the transition linewidth and the voltage dependence of $\lambda_{bg}^{Q}$
are consistent with the quantum-confined Stark effect \cite{stolarz_11}. 

\begin{figure}[b]
\centering
\includegraphics[trim=0.5cm 1cm 4cm 2cm, clip=true, width=0.975\columnwidth]{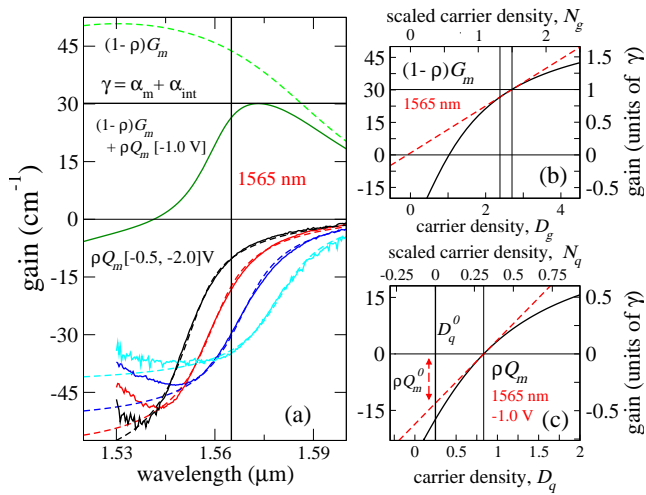} 
\caption{\label{fig5} (a) Fits to the modal absorption at a series of values of the reverse bias as shown.
The modal gain corrected for the finite absorber length, and the net gain in the two-section device for 
a reverse bias of -1.0 V are also shown.  (b) Variation of the modal gain in the two-section 
device with carrier density at the reference wavelength indicated in (a). Vertical lines indicate the 
threshold carrier density in the FP and the threshold carrier density in the two-section device assuming a transparent 
absorber section. (c) Variation of the modal absorption in the two-section device with carrier density. 
The wavelength is the same as in (b) The scaled carrier density is zero in the unsaturated device, and the 
differential absorption is defined at the saturated value. }
\end{figure}

Experimental and fitted absorption curves for four values of the reverse bias voltage are shown in Fig. \ref{fig5} (a). 
In each case, at short wavelengths, the experimental curves show a region of decreasing absorption with decreasing 
wavelength. This non-monotonic dependence is consistent with the measurements of Fig. \ref{fig3} and it 
is not described by our model absorption function. The result is that fits to these absorption curves are only accurate 
for wavelengths above a characteristic cutoff that depends on the applied voltage. In all cases however, one can see 
that the fits are very close to the measured data for wavelengths larger than approximately 1555 nm. From the 
optical spectra of the lower panel of Fig. \ref{fig1} and from Fig. \ref{fig2}, we can see that the peak emission 
wavelength begins to shift towards short wavelengths and approach 1555 nm only for reverse bias voltage values of -1.5 V or 
larger. These values also define the boundary of the region of rapid variation of the peak emission with voltage, which will 
result in poor tunability of the ML spectrum. For this reason we will illustrate the application of our optimization method 
targeting reverse bias voltages between -1.0 V and -1.5 V, and assume that our fits to the absorption are accurate at all 
wavelengths of interest over this voltage range. A table of fitting parameters for the modal absorption function obtained 
for voltages ranging from -0.5 to -1.5 V are shown in table \ref{t2}. Here $D_q^{0}$ is the value of the carrier density 
used to fit the measured [unsaturated] absorption.

\begin{table}[t]
\renewcommand{\arraystretch}{1.5}
\begin{center}
\begin{tabular}{ccccccccc} 	\hline
bias & $\lambda_{bg}^{Q}$ & $\rho Q_{0}$ & $D_q^{0}$  & $\lambda_{0}$ & $\rho Q_{m}^{\scriptscriptstyle 0}$  
& $N_g^{\scriptscriptstyle 0}$ & $g_{g}$ & $\rho g_{q}$  \\ \hline
[V] & [nm] & [$\mbox{cm}^{-1}$] &  & [nm] & [$\gamma$] & &  & \\ \hline
-0.5 & 1558.0 &  22.75 &  0.33 & 1560.0 & 0.38 & 1.36 & 0.79 & 1.39 \\  \hline
-1.0 & 1564.0 &  20.75 &  0.25 & 1565.0 & 0.43 & 1.43 & 0.68 & 1.40  \\  \hline
-1.5 & 1572.0 &  18.25 &  0.18 & 1571.5 & 0.49 & 1.60 & 0.54 & 1.41  \\  \hline
\end{tabular}
\caption{ \label{t2} Fitting and derived parameters obtained for the absorber section as a function of reverse
bias voltage. }
\end{center}
\end{table} 

In order to define our model parameters, we must choose a reference wavelength and a carrier density value
for each device section where the modal gain and absorption functions are linearized. For the gain variable, we 
expand around the carrier density value that corresponds to the threshold gain assuming a transparent absorber 
section: $G_m = G_{m}^{0} + g_{g} ( N_{g} - N_{g}^{\scriptscriptstyle 0} )$, where $G_{m}^{0} = G_m(N_{g}^{\scriptscriptstyle 0}) = 
(1 - \rho)^{-1} \gamma$. For the absorber section, we linearize the absorption function at the saturation value 
of the carrier density, and define the carrier density variable so that the unsaturated modal losses, $Q_{m}^{0}$, 
correspond to zero carrier density in the absorber section. The modal absorption is then given by $Q_m = g_{q} N_{q} - 
Q_{m}^{0}$. We will see that our choice of linearization point for the short section is appropriate given the location 
of the stable mode-locking region far from threshold, where the absorber is strongly saturated.  

Results for a reverse bias voltage of -1.0 V are illustrated in Fig. \ref{fig5} (b) and (c), where 
we have plotted the modal gain and absorption as a function of carrier density at the reference wavelength indicated 
by the vertical line in Fig. \ref{fig5} (a). We will explain how this wavelength was chosen in the next section. The 
differential gain at the threshold carrier density assuming a transparent absorber section is $g_g = 0.68$ or 
$10.8 \mbox{ cm}^{-1}$ in physical units, and the differential absorption at the saturation value is $\rho g_q = 1.4$ or 
$22.4 \mbox{ cm}^{-1}$ (here we have quoted values for the differentials that account for the difference in the device 
section lengths). The calculated differential gain $g_g$ and differential absorption $\rho g_q$ for voltages ranging 
from -0.5 to -1.5 V are shown in table \ref{t2}. Here we have also included the reference wavelength $\lambda_0$, to be 
derived in the next section, the unsaturated losses, $\rho Q_{m}^{0}$, and the threshold carrier density 
$N_g^{\scriptscriptstyle 0}$ as defined above. The differentials are given in dimensionless units defined by the total losses 
$\gamma$.

\section{Delay-differential mode-locked laser model and device optimization method}

The results of the previous section have provided us with a qualitatively accurate picture of the variation of 
the modal gain and absorption in the FP device as functions of carrier density, wavelength and reverse bias voltage. 
In this section we introduce the delay-differential model of the two-section device, and we describe our method
for finding the optimum wavelength location for a spectral filter based on the measured material dispersion. In 
order to test the accuracy of the delay-differential model, our first objective will be to reproduce a portion of the 
phase-space structure shown in Fig. \ref{fig1}, concentrating on the region where the variation of the threshold
current with reverse bias voltage is normal. 

The delay-differential lumped element model that we employ eliminates the spatial dependence of distributed time-domain 
models in favour of a delay-differential equation for the field variable. As we have discussed, this model accounts 
for the large gain and loss and the strong saturation of the absorption that are typical of mode-locked semiconductor 
lasers. It is also suitable for analytic analysis, and it requires minimal computing resources to implement. Although
the model is derived assuming a ring-cavity geometry, previous results have shown qualitative agreement with
distributed time domain simulations of mode-locking in linear two-section geometries with both
quantum well and quantum dot active regions \cite{bandelow_06, rossetti_11}. 

The system of equations is \cite{vladimirov_04, vladimirov_05, raghunathan_13}
\begin{eqnarray}\label{dde_sms}
& \gamma_{\scriptscriptstyle G} \dot{E}(t) = -E(t) + \sqrt{\kappa} R(t - \tau) E(t-\tau) \nonumber \\
& \dot{G} = g_0 - \gamma_r G - \exp -Q (\exp G -1)|E|^2 \\
& \dot{Q} = q_0 - Q - s (1 - \exp-Q)|E|^2 \nonumber
\end{eqnarray}
where 
\begin{displaymath}
R(t) = \exp\left[\frac{1}{2} (1 - i \alpha_g) G - \frac{1}{2} (1 - i \alpha_q) Q - i \phi\right]
\end{displaymath}
and
\begin{displaymath}
\kappa = \exp\left[ - (\alpha_{\mbox{\tiny m}} + \alpha_{\mbox{\tiny int}}) L_c\right].
\end{displaymath}
In the above, $\tau$ is the round-trip time in the cavity, the dynamical variable for the gain is defined as 
$G = \int_{(1-\rho)L_c} dz G_m(z,t)$, and the saturable absorption is described by the variable 
$Q = \int_{\rho L_c} dz Q_m(z,t)$. Here $G_m(z,t)$ and $Q_m(z,t)$ are respectively the modal gain and absorption, which 
are now assumed to be spatially varying. $\gamma_r = \gamma_g/\gamma_q = \tau_q/\tau_g$ is the ratio of the recovery 
times in the gain and absorber sections, and $s = g_q/g_g$ is the ratio of the material differential gain and 
absorption. $\alpha_{g,q}$ are the linewidth enhancement factors in the gain and absorber sections of the device, 
and $\phi$ is the detuning of the gain peak from the nearest cavity mode. With the linearized gain and absorption 
variables defined as in the previous section, the scaled pump current is 
$g_0 = \int_{(1-\rho)L_c} dz [ \gamma_q^{-1} g_g j - (g_g N_g^{\scriptscriptstyle 0} -  G_{m}^{0})]$ and 
the parameter describing the unsaturated absorption is $q_{0} =  \int_{\rho L_c} dz Q_m^{\scriptscriptstyle 0}$.
The parameter $\gamma_{\scriptscriptstyle G}$ determines the bandwidth of the gain medium, which is included  
in these equations through the action of a linear Lorentzian filter \cite{vladimirov_05}.

We have seen that the peak emission wavelength of the FP laser will in general evolve continuously as the drive 
parameters are varied. On the other hand, for each bias voltage value, each wavelength defines a unique set of 
model parameters through the modal gain and absorption functions. In order to reproduce the structure of the phase-space 
of Fig. \ref{fig1}, our approach therefore is to associate a unique wavelength with each point in the phase-space of 
drive parameters. To do this, we calculate the variation of the net gain along branches of steady-state solutions 
of Eqns \ref{dde_sms}. The voltage value and material parameters at each wavelength define a unique branch of 
steady-state solutions in the current density variable. To compare the model and experiment, we choose the branch whose 
wavelength coincides with the peak of the net gain at the current density point of interest. This method provides a 
consistent means to compare the measured phase-space data and the simulation results over the entire parameter space, 
provided the variation of the net gain is well described by our fitted modal gain and absorption functions. To avoid 
unnecessary details, we will take -1.5 V as defining a boundary beyond which fitting errors can lead to disagreement 
between the calculated and measured location of the peak emission wavelength. In practice, one should locate this boundary 
precisely as a function of device current and define the target ML region for voltages inside this boundary. Note that we 
should also in principle linearize our modal absorption function at the carrier density value where the net gain peak and 
the wavelength of the steady-state solution branch in question coincide. However, finding the correct steady-state 
branch in this case requires a much larger numerical effort, as the dispersion of the net gain at any particular 
wavelength will vary as the linearized absorption function is changed. Instead, we linearized the modal absorption 
around the transparency point, which is only approached asymptotically. However, we will see that the absorption is 
already well saturated inside the large stable mode-locking region of Fig. \ref{fig1}. 

To find the correct steady-state solution branches we must first calculate the steady-state solutions of the model. 
Assuming slowly varying fields with respect to $\tau$, we neglect the time derivative term in Eqn \ref{dde_sms} and 
multiply the resulting equation by its complex conjugate to obtain $P(t + \tau) = \kappa \exp[G(t) - Q(t)] P(t)$, 
where $P = |E|^2$ is the optical power. The assumption of slowly varying fields implies that 
$P(t + \tau) = P(t) + \tau \dot{P}(t)$ so that we obtain an ordinary differential equation for the optical power 
\cite{rachinskii_06}
\begin{equation}
\tau \dot{P} = -[1 - \exp(G -  Q + \ln \kappa)]P.
\end{equation}
This equation, together with the equations for the gain and absorption variables in Eqn \ref{dde_sms}, can be regarded 
as defining an improved LSA model that is accurate for large gain and loss. In order to find the steady-state solution 
branches we must specify the values of the carrier recovery times in the gain and absorbing sections of the device. 
We can estimate the carrier recovery time in the gain section from a measurement of the relaxation oscillation 
frequency of the single section FP device. If we make the approximation that $g_g N_g^{0}/\gamma = 1$ for the plain FP 
laser, this quantity is expressed in physical units as
\begin{equation}
\nu_{\mbox{\tiny RO}} = \frac{1}{2\pi} \sqrt{\frac{\gamma_{\mbox{\tiny phtn}} (j_s - 1)}{\tau_g}},
\end{equation}
where $\gamma_{\mbox{\tiny phtn}} = \gamma v_g$ is the photon lifetime in the cavity and $v_g$ is the group velocity. The 
relaxation oscillation frequency can in turn be determined from the intensity noise power spectrum \cite{bitauld_10}. 
At a device current of 40 mA $[j_s \simeq 3]$, we found that the relaxation oscillation frequency is approximately 4 GHz, 
from which we determine that $\gamma \tau_g \simeq 240 $ or $\tau_g = 0.9$ ns. To complete the improved LSA model we must 
fix the value of the carrier lifetime in the absorber section. This quantity is in general smaller than the recovery time in 
the gain section, and it is known to have a strong dependence on the applied voltage. To determine $\tau_q$ for our simulations, 
we assumed that the recovery time varies as $\tau_q = \tau_q^{0}\exp[\mbox{V}/\mbox{V}_0]$  \cite{mulet_06, karin_94} and 
took values of $\tau_q^{0} = 85$ ps and $V_0$ = 3.0 V that provided the best agreement with experiment. These values then 
determine $\tau_q$[-0.5, -1.0, -1.5] V = [72, 61, 52] ps. 

\begin{figure}[hbt]
\centering
\includegraphics[trim=1.5cm 6cm 7cm 2cm, clip=true, width=0.98\columnwidth]{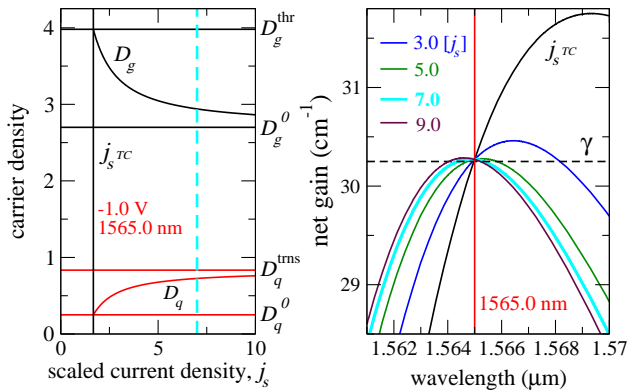}
\caption{\label{fig6} Left: Carrier densities $D_g$ and $D_q$ in both sections of the device calculated along the steady-state 
branch at a wavelength of 1565 nm for a reverse bias voltage of -1.0 V. The solid vertical line is the threshold current density, 
$j_s^{\mbox{\tiny TC}}$, for this branch of solutions. The dashed vertical line is the current density value [$j_s^{*} = 7$] where the 
peak of the net gain along the steady-state branch coincides with the wavelength reference. Solid horizontal lines are the threshold 
carrier density in the gain section, $D_g^{\mbox{\tiny thr}}$, the threshold carrier density in the gain section, assuming a transparent 
absorber section, $D_q^{\mbox{\tiny thr}}$, and the saturated, $D_q^{\mbox{\tiny trns}}$, and unsaturated, $D_q^{0}$, carrier densities in the 
absorber section. Right: Variation of the net gain along the steady-state branch of the figure on the left for a series of current 
density values as indicated.}
\end{figure}

Results showing the variation of the carrier density in both sections of the device along a steady-state solution 
branch located at a wavelength of 1565 nm for a reverse bias voltage of -1.0 V are shown in the left panel of Fig. 
\ref{fig6}. In the figure the threshold current density, $j_s^{\mbox{\tiny TC}}$, for this branch of solutions is indicated.
Horizontal lines indicate the threshold carrier density in the gain section, $D_g^{\mbox{\tiny thr}}$, the threshold 
carrier density in the gain section, assuming a transparent absorber section, $D_g^{0}$, and the saturated, 
$D_q^{\mbox{\tiny trns}}$, and unsaturated, $D_q^{0}$, carrier densities in the absorber section. The wavelength reference 
was determined by requiring that the peak of the net gain along the steady-state branch coincided with the wavelength 
reference at a scaled current density $j_s^{*} = 7$ [83 mA]. This value was chosen as it lies near the center of the 
region where we found the highest quality mode-locking in the FP device. The variation of the net gain along the 
branch of steady-state solutions is shown in right panel of the figure. One can see that the peak of the net gain is
located relatively far [5 nm] from the wavelength reference at threshold, with an excess of the net gain of approximately
1.5 $\mbox{cm}^{-1}$. As the current density is increased, the peak of the net gain shifts rapidly towards the wavelength 
reference, where the net gain is clamped at the value of the total losses $\gamma$. Because the current density region 
we are targeting here is located far above threshold, we find that the variation of the net gain is minimal over a 
large range of current density values near the target. This is due to the fact that the absorption is already 
well saturated in this region, and it justifies our expanding the absorption function around the saturation value 
of the carrier density. These results also suggest that we can be confident that our choice of model parameters at the 
wavelength reference will be appropriate for current density values in excess of $j_s \simeq 5$ 
[59 mA].

\begin{figure}[t]
\centering
\includegraphics[trim=2.5cm 1.5cm 2.5cm 1.5cm, clip=true, width=0.98\columnwidth]{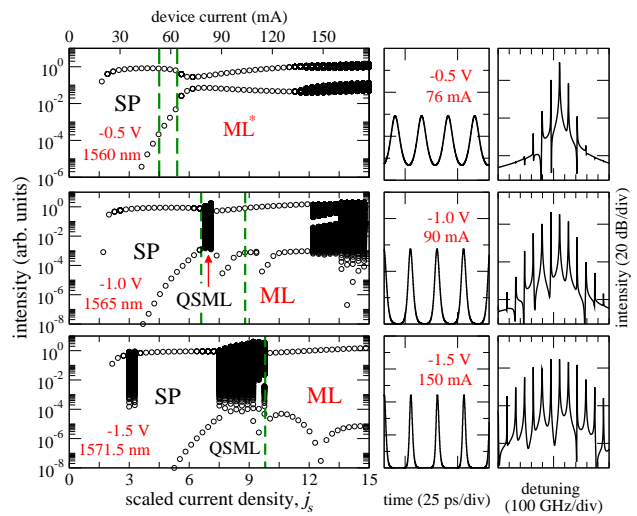}
\caption{\label{fig7} Left panels: Numerical bifurcation diagrams calculated at three reverse bias values
with model parameters defined at the reference wavelengths as indicated. Dashed vertical lines indicate 
mode-locking stability boundaries calculated using the generalized New stability criterion of Ref. 
\cite{vladimirov_05}. Right panels: Intensity time traces, and corresponding optical spectra at values of the 
current density located in the mode-locking regions as indicated.}
\end{figure}

We now present numerical bifurcation diagrams calculated using the delay-differential model and compare the
results with the experimental measurements from the plain FP laser of Fig. \ref{fig1}. For these simulations 
the value of $\gamma_{\scriptscriptstyle G}^{-1}$, which determines the bandwidth of the Lorentzian filter in Eqns 
\ref{dde_sms}, was set as 0.15 ps. This value corresponds to a bandwidth of approximately 80 cavity modes, and 
it was chosen so that the curvature of the modal gain function at the gain peak was accurately reproduced by the 
Lorentzian filter. Further parameters were $\alpha_g = \alpha_q = 2.6$, and $\phi = 0$. The value of $\alpha$ 
chosen is consistent with our previous work modeling optical injection and feedback experiments in devices 
with similar material composition \cite{osborne_12}. 

Numerical bifurcation diagrams calculated for reverse bias values ranging from -0.5 V to -1.5 V are shown in Fig.
\ref{fig7}. In each case the value of the wavelength reference is indicated. One can see that the variation
of the reference wavelength with bias voltage from 1560 nm at -0.5 V [$j_s^{*} = 5$] to 1565 nm at -1.0 V [$j_s^{*} = 7$] 
is in broad agreement with the measured variation of the peak emission wavelength visible in the optical spectra 
shown in Fig. \ref{fig1}. The further red-shift in the reference wavelength to 1571.5 nm that we find for a bias 
voltage of -1.5 V [$j_s^{*} = 9$] is not in agreement with the measured data for -1.6 V, as this voltage is 
inside the region where the anomalous dispersion of the material absorption comes into play. However, we will see 
that the dynamical states of the laser are qualitatively reproduced in this voltage region. If we first compare 
the simulated device dynamics with measured data at -0.5 V, we find SPs at threshold, followed by a transition to 
a dynamical state where the intensity is weakly modulated at a frequency close to the round trip time in the cavity. 
Although SPs at threshold were not resolved in our measurements at this voltage, for larger currents we recall that 
the corresponding measured behavior was incomplete mode-locking, where a large number of FP modes were above threshold 
but the intensity modulation at the round trip frequency was also weak. For comparison, we have included a 
simulated intensity time trace and corresponding optical spectrum at a scaled current density of $j_s = 6.5$ [76 mA]. 
One can see that the optical spectrum is almost single mode for these parameters. We have therefore referred to these 
states as ``starred'' mode-locked states of the laser, [$\mbox{ML}^{*}$], as the intensity modulation at the round-trip 
frequency is very weak in this region. The tentative correspondence between these simulations and our measurements is 
an indication that the delay-differential model with calibrated parameters can capture subtle physical variations that 
were found across the ML region in our experiment. 

At a reverse bias of -1.0 V, we see that the SP region at threshold increases in size, and we find a narrow region 
of QSML before entering a region of well developed mode-locking near a device current of 90 mA. This narrow region 
of QSML is also evident in our experimental measurements for larger reverse bias values. The ML region extends as 
far as 140 mA, where mode-locking becomes unstable. When compared to our experimental measurements, we can see that 
the ML region here opens at a larger value of the device current, and that it extends over a larger current range 
in the simulation. However, one should note that the incomplete ML region near -0.5 V was recorded 
as a CW region in our measurements, which indicates that there may be a larger region of incomplete ML surrounding the 
labeled ML region in Fig. \ref{fig1}. If we examine the simulations for -1.5 V in the lower left panel of Fig. 
\ref{fig7}, one can see that the general trends that were observed in our measurements are well reproduced by the 
simulations. These include the increasing size of the QSML region, and the rapid expansion of the ML region with 
increasing reverse bias. One can see also that the pulse duration becomes progressively shorter as the reverse bias 
is increased. Although the numerical pulse duration at -1.0 V [2.7 ps] is considerably longer than the experimental 
value, the pulse duration is 1.6 ps at -1.5 V, which is in broad agreement with measured data.  

Dashed vertical lines in the bifurcation diagrams of Fig. \ref{fig7} indicate regions of stable mode-locking 
calculated using the generalized New stability criterion [GNSC] for the DDE model formulated in Ref. 
\cite{vladimirov_05}. The New stability criterion for mode-locked states requires that the net gain must be 
negative at the beginning and end of the slow-stage of the dynamics between pulses when the intensity is small 
\cite{new_74}. Although the agreement with the GNSC is poor for smaller reverse bias voltages, at larger reverse 
bias, where the pulses are well developed, we find this criterion gives a very good estimate for both stability 
boundaries [the upper ML stability boundary at -1.5 V near $j_s = 16 $ is not shown in this figure]. The utility 
of these semi-analytic results for the ML stability boundaries obtained from Eqns \ref{dde_sms} highlight an attractive 
feature of the DDE model, but they also demonstrate that the DDE model has a much richer dynamics and a much larger 
stable ML region than the GNSC would in general suggest.  

The results presented in Fig. \ref{fig7} demonstrate that the delay-differential model with experimentally 
calibrated parameters can provide reasonably good qualitative agreement with experimental data. Overall, the 
model seems to overestimate the drive current required to reach the ML region at a given voltage, and it also 
overestimates the size of the various regions of dynamics in both the voltage and device current variables.
It is clear however that the region of simulated high quality mode-locking evolves from a region of incomplete 
mode-locking as the reverse bias voltage is increased, which means that we can judge if a simulated ML region is 
likely to vanish for a small change in the model parameters. 

With these qualifications in mind, we now consider the problem of finding the optimal wavelength location for a 
spectral filter in order to maximize the stable mode-locking region and tunability of the device. If we imagine a 
relatively narrow-band filter, and a quasi-static picture of the dynamics, then the current and voltage range over 
which we can expect stable mode-locking at the desired wavelength will be determined by the variation of the net gain 
along the corresponding steady-state branches at the wavelength location of the spectral filter. To avoid transfer of 
the optical power to adjacent cavity modes, we require that the excess of the net gain along a steady-state branch at 
the filter location cannot exceed the filter selectivity at detuned wavelengths. This condition must be then be 
satisfied over a continuous region in the current and voltage parameters, accounting for the fact that each voltage 
determines a different steady-state branch at the wavelength of the spectral filter. 

\begin{figure}
\centering
\includegraphics[trim=1.5cm 6cm 9cm 2cm, clip=true, width=0.98\columnwidth]{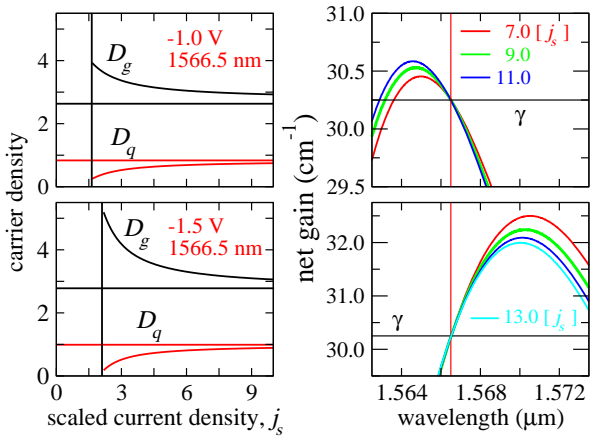}
\caption{\label{fig8} Carrier densities calculated along the steady-state branches at 1566.5 nm 
(left panels). Variation of the net gain for a series of current density values on the same branch as 
indicated (right panels).}
\end{figure}

\begin{figure}
\centering
\includegraphics[trim=1cm 6cm 8cm 1cm, clip=true, width=0.98\columnwidth]{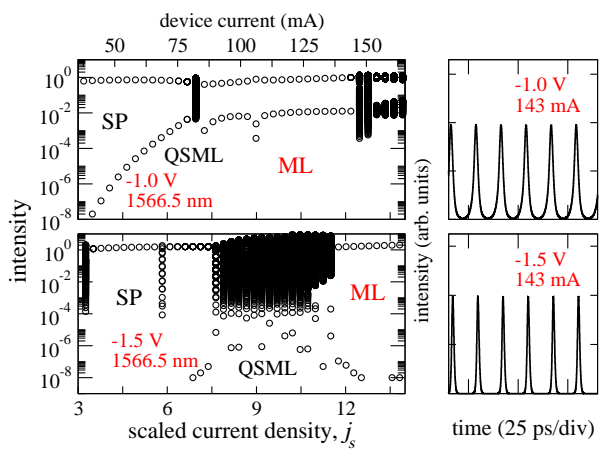}
\caption{\label{fig9} Bifurcation diagrams, time traces, and optical spectra. Top panels: -1.0 V, and lower 
panels -1.5 V. The wavelength that determined the model parameters is 1566.5 nm, and the pump parameter value
for each of the time traces was $j_s = 12$. }
\end{figure}

An experimentally calibrated example illustrating the application of this method is shown in Figs \ref{fig8} and 
\ref{fig9}, where we have found that a filter located at 1566.5 nm will lead to a large and continuous region of 
high quality mode-locking near 140 mA for voltages between -1.0 and -1.5 V. Left panels in Fig. \ref{fig8} show 
the carrier density along the steady-state solution branches for the two voltages, while the right panels show the 
variation of the net gain along these branches for $j_s$ between 7 [83 mA] and 13 [154 mA]. Notice how the wavelength 
reference here is located closer to the peak of the net gain at -1.0 V, which leads to a larger excess of the net gain 
along the branch at -1.5 V. This was necessary in order to ensure overlapping ML regions in the phase-space
as shown in the left panels of Fig. \ref{fig9}. Time traces in the right panels of Fig. \ref{fig9} indicate well 
developed pulses over the ML region. Results in Fig. \ref{fig8} indicate that a spectral filter will require a 
minimum selectivity of ca. 3 $\mbox{ cm}^{-1}$ to ensure stable emission at the desired wavelength over this parameter 
region. Although this value of the filter selectivity exceeds the measured value from the FP device of Fig. \ref{fig3}, 
we note that this device was based on a low-density grating structure that was defined by etched features in the laser 
ridge waveguide. Numerical simulations of distributed feedback structures that define comb laser spectra indicate that 
more advanced grating designs can readily achieve this level of selectivity \cite{bitauld_11b, obrien_14}.

The results of Fig. \ref{fig9} indicate that a grating spectral filter centered at a wavelength of 1566.5 nm 
will allow for tuning of the ML state of the device over a wide range in the drive current and voltage parameters. 
Although the overall device length and absorber section length of the device of Fig. \ref{fig3} were not identical to 
the FP laser we considered in order to illustrate the application of the method, the material composition was identical. 
It is notable therefore that the result for the optimal location of the grating filter is some 20 nm from the 
designed wavelength location of the device of Fig. \ref{fig3}. We note also that the optimized wavelength location
is more than 12 nm from the position of the gain peak at threshold in the FP device. 

Because the agreement between the simulated and measured phase-space data for the FP device was only qualitative, our 
results for the optimized wavelength location and the predicted dynamics are approximate. In particular, we did not account 
explicitly for any additional dispersion or change in modal thresholds that a grating filter might introduce in the cavity. 
Numerical results have shown that the grating induced dispersion can become much larger than the underlying material 
dispersion in devices with strongly scattering filters \cite{bitauld_11b}. However, it is also possible in general to 
correct for this in narrow-band examples, which are the focus of our interest here. We also note that the comparison between 
the simulations and experiment seemed to indicate the presence of some systematic errors, where the onset of mode-locking 
and the width of the ML region in the current variable was overestimated. It would be interesting to establish if these 
errors could be eliminated with a more accurate model of the semiconductor susceptibility. 

We note also that although the DDE model is a large signal model of the mode-locked laser, our method was
based on steady-state solutions of the model and is therefore small signal in nature. We justified this limitation
on the grounds that we are interested primarily in mode-locked devices with relatively narrow-band spectra. It
would however be interesting to attempt to extend the method in order that it could be applied to devices designed 
to support wide bandwidth combs, with the potential to generate short pulses. Under these circumstances, the 
spectrum could be strongly influenced by dynamical effects such as self-phase modulation and the efficiency 
of the absorber saturation in a mode-locked state \cite{bischoff_95, javaloyes_10}. 

Although we believe that these results demonstrate the utility of the DDE model, our work has also highlighted a 
number of its limitations. The DDE model is derived assuming a linear Lorentzian spectral filter that corresponds to 
the gain bandwidth of the semiconductor. However, we obtained unphysical results if we attempted to adjust the filter 
bandwidth parameter to values that might describe a device such as in Fig. \ref{fig3}. In addition, we found that 
adjusting the nominal value of the detuning of the gain peak from the nearest cavity mode through the $\phi$ 
parameter had a significant effect on the simulated phase-space. Given that the gain bandwidth was of order 80 cavity 
modes, this is also an unphysical dependence of the model. Similarly, we found a strong dependence on the alpha factor 
values chosen for both sections of the device. Although this dependence is in some respects to be expected, we were 
unable to interpret our results in a consistent way. We therefore set the alpha parameters to be equal for both 
sections, and we leave a detailed exploration of the combined roles of the alpha factors and $\phi$ parameter for 
future work. Overall, it seems that further development of the DDE or a similar model is both necessary and desirable 
given its obvious strengths. We note that some progress towards generalizing the action of the spectral filter to 
describe waveguide dispersion in real devices has been already been made \cite{heuck_10}. 

\section{Conclusions}

In conclusion, we have presented a method for determining the optimal wavelength location for a grating spectral 
filter in a two-section semiconductor mode-locked laser. The goal of this method was to maximise the region of
stable mode-locking in the phase-space of drive parameters of the device. We accounted for material
dispersion in the two sections of the laser using a simple analytic model for the semiconductor susceptibility,
which provided good agreement with measured data over the parameter region of interest. Our dynamical
simulations were based on a delay-differential model of the device, which we found could qualitatively 
describe the structure of a measured dynamical phase-space when calibrated with experimental parameters. By 
considering the variation of the net gain along steady-state solutions of the model, we were able to optimize 
the location of the spectral filter with respect to regions of stable mode-locking in the device. Our results
indicate that it will be possible to obtain a large region of stable mode-locking in devices with practical
values of the grating selectivity. 

The authors acknowledge financial support from Science Foundation Ireland under grant SFI13/IF/I2785. 


\end{document}